\documentclass[manuscript]{aastex}
\usepackage{graphicx}
\usepackage{amssymb}

\begin{document}

\newcommand{\feii}{\ion{Fe}{2}}
\newcommand{\hi}{\ion{H}{1}}
\newcommand{\NIii}{\ion{Ni}{2}}
\newcommand{\Sii}{\ion{S}{2}}
\newcommand{\Siii}{\ion{S}{3}}

\slugcomment{To appear in Ap. J.}

\title{Empirical determination of Einstein $A-$coefficient ratios of bright [\feii] lines\thanks{Based on observations collected with X-shooter at the Very Large
Telescope on Cerro Paranal (Chile), operated by the European Southern
Observatory (ESO). Program ID: 092.C-0058(A).}}

\author{T. Giannini\altaffilmark{1}, S. Antoniucci\altaffilmark{1}, B. Nisini\altaffilmark{1}, D. Lorenzetti\altaffilmark{1},J. M. Alcal\'a\altaffilmark{2}, F. Bacciotti\altaffilmark{3},\\ R. Bonito\altaffilmark{4,5}, L. Podio\altaffilmark{3}, B. Stelzer\altaffilmark{4}
}
\altaffiltext{1}{INAF-Osservatorio Astronomico di Roma, via Frascati 33, I-00040 Monte Porzio Catone, Italy}
\altaffiltext{2}{INAF-Osservatorio Astronomico di Capodimonte, via Moiariello 16, I-80131 Napoli, Italy} 
\altaffiltext{3}{INAF-Osservatorio Astrofisico di Arcetri, Largo E. Fermi 5, I-50125 Firenze, Italy} 
%$^4$ Department of Physics, Western Michigan University, Kalamazoo, MI 49008, USA\\
\altaffiltext{4}{INAF-Osservatorio Astronomico di Palermo, Piazza del Parlamento 1, I-90134 Palermo, Italy}
\altaffiltext{5}{Dipartimento di Fisica e Chimica, Universit\'a di Palermo, Piazza del Parlamento 1, I-90134 Palermo, Italy}
%\altaffiltext{6}{UJF-Grenoble 1 / CNRS-INSU, Institut de Planetologie et d'Astrophysique de Grenoble (IPAG) UMR 5274, Grenoble, F-38041, France}
%\altaffiltext{7}{Institut f\"{u}r Astronomie und Astrophysik, Kepler Center for Astro and Particle Physics, Eberhard Karls Universit\"{a}t, 72076 T\"{u}bingen, Germany}

\begin{abstract}
The Einstein spontaneous rates ($A-$coefficients) of Fe$^{+}$ lines have been computed by several authors, with results that differ from each other up to 40\%. Consequently, models for line emissivities  suffer from uncertainties which in turn affect the determination of the physical conditions at the base of line excitation.
We provide an empirical determination of the $A-$coefficient ratios of bright [\feii] lines, which would represent both a valid benchmark for theoretical computations and a reference for the physical interpretation of the observed lines.
With the ESO-VLT X-shooter instrument between 3\,000 \AA\, and 24\,700 \AA\,, we obtained a spectrum of the bright Herbig-Haro object HH\,1. We detect around 100 [\feii] lines, some of which with a signal-to-noise ratio $\ge$ 100. Among these latter, we selected those emitted by the same level, whose de-reddened intensity ratio is a direct function of the Einstein $A-$coefficient ratios. From the same X-shooter spectrum, we got an accurate estimate of the extinction toward HH\,1 through intensity ratios of atomic species, \hi\, recombination lines and H$_2$ ro-vibrational transitions. We  provide seven reliable $A-$coefficient ratios between bright [\feii] lines, which are compared with the literature determinations. In particular, the  $A-$coefficient ratios involving the brightest near-infrared lines ($\lambda$12570/$\lambda$16440 and $\lambda$13209/$\lambda$16440) are better in agreement with the predictions by Quinet et al. (1996) Relativistic Hartree-Fock model. However, none of the theoretical models predicts $A-$coefficient ratios in agreement with \emph{all} our determinations. We also show that literature data of near-infrared intensity ratios better agree with our determinations than with theoretical expectations.

\end{abstract}


\section{Introduction}{\label{sec:sec1}
Spectra of nebular environments are commonly characterized by forbidden atomic lines in emission. Usually, the 
brightest lines are those coming from the fine structure, ground-state levels of the most abundant species, such as oxygen, nitrogen, and sulphur. Although bright, these lines are few in number and sensitive to 
specific excitation conditions, being therefore unable to trace gradients in the physical parameters inside the gas (e.g. Bacciotti \& Eisl\"{o}ffel 1999; Giannini et al. 2013). Moreover, diagnostic tools based on flux ratios between lines of different species require to  assume the relative elemental 
abundances, which in turn imply an uncertainty on the derived physical parameters that can exceed 40\% (Podio et al. 2006). All these limitations can be circumvented by using flux ratios of lines from the same species, which cover a wide range of excitation conditions and are also valuable to probe the extinction value. In this sense, UV to near-infrared [\feii] lines, have been recently proven as being a very powerful diagnostic tool (Giannini et al. 2013). The main limitation of iron diagnostics derives from the computational method adopted to describe the quite complex atomic system of iron, from which the derivation of atomic parameters (both radiative and collisional) depends. By combining the uncertainties on $A-$coefficients, collisional coefficients and their propagation on the level populations, Bautista et al. (2013) have shown that the uncertainties on the line emissivity can be larger than 60\%. These uncertainties imply a poor determination of the physical parameters: for example a 
difference of 30\% in the $A-$coefficients in the [\feii]$\lambda$12570/$\lambda$16440 flux ratio, which is commonly used to estimate the reddening, gives an uncertainty on A$_V$ of about 3 magnitudes.\\
In addition to a number of theoretical efforts (Nussbaumer \& Storey 1988; Quinet et al. 1996; Deb \& Hibbert 2011), an attempt to empirically derive the rates for spontaneous emission of bright [\feii] lines,  has been done by Smith \& Hartigan (2006) by means of observations on the P Cygni's nebula. This method, originally proposed by Gredel (1994), considers that the intensity ratio of optically thin lines coming from the same upper level does not depend on the level population (and hence on temperature and density of the emitting gas and/or source variability), but rather is simply $I_{ij}/I_{ik} = A_{ij} \nu_{ij}/A_{ik} \nu_{ik}$, where A$_{ij}$, A$_{ik}$ and $\nu_{ij}$, $\nu_{ik}$ are the Einstein $A-$coefficients and the frequencies of the \emph{i}$\rightarrow$\emph{j}, \emph{i}$\rightarrow$\emph{k} lines emitted by the \emph{i}-level.
This observational approach has the definite advantage of providing a direct measure of the $A-$coefficient ratios, but it can be applied only if the extinction for which the observed line intensities are de-reddened is known with great accuracy. In the case of P Cygni's nebula, the assumed extinction (A$_V$ =1.89, Lamers et al. 1983) gives, for example, an intrinsic $\lambda$12570/$\lambda$16440  $A-$coefficient ratio of 1.13, which is definitively higher than all the theoretical computations provided in the literature (roughly ranging between 0.79  and 1.04). In a previous work (Giannini et al. 2008), we compared this value (together with all the other available in the literature at that time) with a number of observations, but we did not find a consistency between observational data and theoretical or empirically derived predictions. \\
In the present work we want to revisit this issue and derive some [\feii] $A-$coefficient ratios by means of observations specifically planned to this aim.
We targeted a well-known and very bright object, namely the bow-show at the edge of the HH\,1 protostellar jet (Raga et al. 2011 and references therein, hereinafter HH\,1). This source is extremely rich in emission lines from different species covering a wide spectral range from the UV to the near-infrared (Solf et al. 1988, Giannini et al, 2014, in prep.), which allow one to derive a precise estimate of the reddening by means of several independent tracers. The target was observed exploiting the unique capabilities of sensitivity and simultaneous spectral coverage of the X-shooter spectrograph at the ESO Very Large Telescope, circumstances that allow us both to get an exceptionally high signal-to-noise ratio (S/N) on the bright lines and to partially overcome possible intercalibration problems among different spectral segments.\\   
The paper is organized as follows: in Sect.\,\ref{sec:sec2} we present the X-shooter observations and the data reduction; in Sect.\,\ref{sec:sec3.1} independent A$_V$ determinations are provided, while in Sect.\,\ref{sec:sec3.2} and Sect.\,\ref{sec:sec3.3} we comment on the $A-$coefficient ratios obtained for bright and valuable diagnostic Fe$^+$ lines. A summary is then given in Sect.\,\ref{sec:sec4}.

\section{Observations and data reduction}{\label{sec:sec2}
We collected seven X-shooter (Vernet et al. 2011) observations of HH\,1 ($\alpha_{J2000.0}$= 05$^{h}$ 36$^{m}$ 20${\fs}$27, $\delta_{J2000.0}$= $-$06$^{\circ}$ 45$^{\prime}$ 04${\farcs}$84, distance of 414 pc), in the period November 1 2013 $-$ January 2 2014, each integrated about 30 min on-source. Since the object (9$^{\prime\prime} \times 12^{\prime\prime}$), fills the 11$^{\prime\prime}$ slit along the spatial direction, we acquired separate sky-exposures integrated as much time as on-source.
The slit, aligned with the direction of the bow-shock (position angle of 129$^{\circ}$), was set to achieve a resolution power of 9\,900,  18\,200 and 7\,780 for the UVB (3\,000 - 5\,900 \AA), VIS (5\,450 - 10\,200 \AA) and NIR arm (9\,900-24\,700 \AA), respectively (slit widths : 0${\farcs}$5,  0${\farcs}$4,  0${\farcs}$6). The pixel scale is 0${\farcs}$16 for the UVB and VIS arm, and 0${\farcs}$21 for the NIR arm. The data reduction was accomplished independently for each arm using the X-shooter pipeline v.2.2.0 (Modigliani et al. 2010),  which provides 2-dimensional spectra, both flux and wavelength calibrated. Post-pipeline procedures were then applied by using routines within the {\it IRAF} package to subtract the sky-exposures, correct the observed radial
velocities for all the motions of the Earth with respect to the direction of the target, combine the seven observations (after having verified that in the single exposures none of the brightest [\feii] lines is contaminated by telluric features), and extract the final 1-dimensional spectra. 
In addition, each NIR image was divided by the spectrum of a telluric standard star taken immediately after the target observation, which was corrected for both the stellar continuum shape and the intrinsic absorption features (hydrogen recombination lines).  
The goodness of the intercalibration
between adjacent arms was checked by comparing fluxes of lines present in the overlapping portions of the spectrum, since this lacks a bright continuum emission. While several lines are present in both the  UVB and VIS spectra, with fluxes that agree within few percent, only one line (\hi\,$\lambda$10048\AA) is detected both in the VIS and NIR arm, being the NIR flux about 40\,\% higher. We
applied this intercalibration factor to the NIR spectrum to re-align it to the other two segments. However, to minimize any possible source of uncertainty, we have avoided to determine $A-$coefficient ratios of [\feii] lines lying in different arms. Also, to derive the extinction value (Sect.\ref{sec:sec3}), we used preferentially ratios of lines lying in the same arm or in the UVB or VIS arms. \\

\section{Results and analysis}{\label{sec:sec3}
The HH\,1 spectrum results exceptionally rich in emission lines, with more than 500 detections of fine structure atomic lines, hydrogen and helium recombination lines and H$_2$ ro-vibrational lines (Giannini et al., in preparation); in particular, around 100 lines come from [\feii] transitions. Those used in this paper to derive the $A-$coefficient ratios have all S/N $\ga$ 100 (see Table\,\ref{tab:tab1} and Figure\,\ref{fig:fig1}). The line fluxes are computed by integrating the emission below the line profile, while the errors are obtained by multiplying the noise (rms) at line base times the full-width at half-maximum (FWHM) of the line profile. 

\subsection{A$_V$ determination}{\label{sec:sec3.1}

As anticipated in Sect.\,\ref{sec:sec1}, a reliable observational estimate of the Einstein  $A-$coefficient ratios can be obtained only provided that the extinction toward the target is determined with a very good accuracy. A powerful way to achieve this is to using pairs of optically thin transitions for which the atomic parameters are accurately known and that come from the same upper level, since in this case the difference between the observed and the theoretical flux ratio is a function only of the extinction amount. 
In the case of HH\,1, thanks to the quite large number of bright lines detected in the spectrum, independent extinction estimates can be derived from a number of line ratios. We consider
hydrogen recombination lines, atomic fine structure lines and H$_2$ ro-vibrational lines. The atomic data ($A-$coefficients and vacuum frequencies) are taken from 'The atomic line list', v2.04, http://www.pa.uky.edu/~peter/atomic/ and from the 'NIST' database, http://www.nist.gov/pml/data/asd.cfm, while H$_2$ frequencies and $A-$coefficients are from Dabrowski (1984) and Wolniewicz et al. (1998), respectively. To minimize the uncertainties, we consider ratios of un-blended lines detected with a S/N $>$ 30. \\
% We also discard all the lines which result blended with lines from other species or contaminated by artifacts in the spectral images. 
 %The considered ratios are listed in Table\,\ref{tab:tab1}, where we also indicate the correspondent A$_V$ determination, which is determined 
Ratios of \hi\, lines coming from the same upper level involve different series, which in our case are the Balmer, Paschen and Brackett series. Given the optimal alignment between the UVB
 and VIS arms, we have considered only ratios of un-blended lines of the Balmer and Paschen series in these two arms (i.e. Balmer lines with n$_{up} \ge$ 4 and Paschen lines with n$_{up} \ge$ 7), with the exception of few ratios involving very bright NIR lines (namely Pa$\delta$/H7; Pa$\gamma$/H$\delta$; Pa$\beta$/H$\gamma$; Br$\gamma$/Pa$\delta$). Lines that appear contaminated by artifacts or sky line residuals in the spectral images are also discarded.
In Figure \,\ref{fig:fig2} we plot the Balmer and Paschen decrements (upper and middle panels) as a function of the upper quantum number. Their distribution is well fitted with a case B (Hummer \& Storey 1987) distribution at 
3\,000\,$\le T \le$\,20\,000 K and low density ($n$=10$^4$ cm$^{-3}$), which ensures the lines to be optically thin. 
To estimate the extinction, we apply the reddening law of Cardelli et al. (1989), after having checked that very marginal differences are found if the Rieke \& Lebosfky (1985) or the Weingartner 
\& Draine (2001) laws are applied. Similarly, the possible effect of a patchy extinction, which should provide a change in the extinction law, is likely negligible. If it were the case, indeed, lines at shorter wavelengths would be more opaque, and therefore a weak trend of the green points to be more above the red ones (as n$_{up}$ increases) should be recognizable.  Also, the ratio of total to selective extinction (R$_V$) was assumed to vary between 3 and 5.\\
The A$_V$ estimate was obtained by averaging each independent determination (twelve lines ratios). We get A$_V$=0.3$\pm$0.1 mag and  A$_V$=0.6$\pm$0.2 mag, for R$_V$=3 and R$_V$=5, respectively. As example, we plot 
in the bottom panel of Figure\,\ref{fig:fig2} the observed (black) and the theoretical ratios (red) along with the observed ratios (green) de-reddened for A$_V$=0.3$\pm$0.1 mag, R$_V$=3.\\
Lower A$_V$ values are derived from line ratios of fine structure lines. In particular we use ratios of [\NIii] ($\lambda$4629/$\lambda$7258 and $\lambda$7380/$\lambda$19393), [\Sii] ($\lambda$4070/$\lambda$10289 and $\lambda$4077/$\lambda$10339) and [\Siii] ($\lambda$9071/$\lambda$9533), which on average give A$_V$=0.15$\pm$0.15 mag for R$_V$=3 (A$_V$=0.30$\pm$0.15 mag for R$_V$=5). Finally, ratios of H$_2$ ro-vibrational lines (2-0Q(i+2)/2-0O(i), with i between 1 and 4), give A$_V$ $\sim$ 0 mag. 
These three determinations, although similar to each other, may indicate that lines from different species likely originate in different areas within the shock, where extinction undergoes a small gradient. 
Since we are unable to determine from which of these areas [\feii] lines originate, we assume assume a range of A$_V$ from 0.0 mag to a conservative value of 0.8 mag.

\subsection{$A-$coefficient ratios of [\feii] lines}{\label{sec:sec3.2}

In Table\,\ref{tab:tab2} we report the theoretical estimates of seven $A-$coefficient ratios  derived by
different authors under different approximations for the Fe$^+$ atomic model (see caption of Table\,\ref{tab:tab2} for references) and the observational determination by Smith \& Hartigan (SH, 2006). In the seventh column we give the  $A-$coefficient ratios derived in the present work. These are presented as a range of values, obtained by combining as uncertainties both the flux errors of Table\,\ref{tab:tab1} and the two extreme extinction estimates of 0.0 mag and 0.8 mag derived in Sect\,\ref{sec:sec3.1}. The last column reports the
intrinsic intensity ratios computed on the base of the derived $A-$coefficient ratios. Notably, none of the literature determination is in agreement with {\it all} our values. 
In particular, for $\lambda$12570/$\lambda$16440 and 
$\lambda$13209/$\lambda$16440 $A-$coefficient ratios we find values close to the Q-HFR predictions, while our determination for $\lambda$15339/$\lambda$16773 is in between the theoretical 
predictions and the SH empirical derivation. A large spread is found between theoretical predictions for $\lambda$7639/$\lambda$8619, which are however all above our derived range. Finally, we 
find  a marginal agreement with the Q-SST, Q-HFR  and DB determinations for $\lambda$7174/$\lambda$7390 and $\lambda$4289/$\lambda$4360 $A-$coefficient ratios, while our range for $\lambda$15995/$\lambda$17116  is in agreement with all the theoretical models and keeps out only the SH determination.\\

\subsection{Comparison with observations}{\label{sec:sec3.3}
The intrinsic intensity ratios of the first two pairs of lines, namely I($\lambda$12570)/I($\lambda$16440) and I($\lambda$13209)/I($\lambda$16440), are of great interest since they involve lines commonly observed in the spectra of nebular environments and that are often used to derive the extinction toward the target. In a previous work
(Giannini et al. 2008, Figure 5) we have already shown how the theoretical intensity ratios are unable to fit the observations of a number of
Herbig-Haro objects where these lines were observed with a high S/N. Here we reproduce the same figure (Figure\,\ref{fig:fig3}) but including both new observational determinations and up-to-date
theoretical determinations. These are depicted as black curves that start at the expected value for A$_V$ = 0 mag and then follow the reddening law of Cardelli et al. (1988) up to A$_V$ $\sim$ 20 mag. Notably, no appreciable differences are found if other extinction laws are adopted (e.g. Rieke \& Lebofsky 1985) also given the short wavelength range taken into consideration. The same reddening law is depicted in red
for the starting points (those for A$_V$\,=\,0 mag) derived in this work, being the open and filled circle the points corresponding to an A$_V$ toward HH\,1 of 0.0 and 0.8 mag, respectively.
By examining the plot several conclusions can be drawn: {\it i}) all the theoretically derived intrinsic intensity ratios are not consistent with the large majority of the data points, irrespective of the extinction values; {\it ii}) the SH curve roughly superposes to that derived in the present work, although starting from a  different A$_V$\,=\,0 mag point. From one side, this result reinforces the goodness of the empirical approach adopted both in the SH and in the present work, but, from the other side, put into evidence that the empirical method is powerful only provided that  
the extinction toward the target is accurately known. In particular, we hypothesize that the extinction value ($\sim$ 2 mag) for which the P Cygni data points have been de-reddened is too high. Indeed, if the SH intrinsic ratio were assumed (I($\lambda$13209)/I($\lambda$16440)\,=\,0.39; I($\lambda$12570)/I($\lambda$16440)\,=\,1.49) the extinction toward HH\,1 would be 
around 2-3 mag, i.e. not in agreement with our estimates; \it iii)} despite the overall better agreement of our predictions with the observational data, a significant number of points, and in particular those associated with low extinction values, tend to remain to the right of our theoretical curve. Although small, e.g. a decrease of less than 10\,\% of I($\lambda$13209)/I($\lambda$16440), would be sufficient to shift the observed data on our theoretical curve, such effect appears to by systematic, and therefore deserves a deeper investigation, in particular as far as possible instrumental effects is concerned.  
In this respect, first we note that, at intermediate spectral resolution, three 
telluric features at $\lambda_{air}$\,=\, 13207.3\,\AA, 16437.7\,\AA \,and 16443.1\,\AA\, (Oliva \& Origlia 1992) may shift into or away the [\feii] lines depending on the terrestrial motion around the Sun at the time of the observation. More important, these telluric features blend fully with the [\feii] lines if the spectral resolution is less than few thousands. Second, flux intercalibration between different spectral segments may have a role, especially for long-lasting observations during which the atmospheric conditions may highly change. Indeed, in Figure\, \ref{fig:fig3}, the largest deviations are associated with low-resolution data taken with spectrographs, such as SofI at ESO-NTT or ISAAC at ESO-VLT, 
which do not cover the three line wavelengths simultaneously. Conversely, the X-shooter observations,
which are taken in one shot and with $R$ larger than 10\,000, are more consistent with our theoretical curve (although with few exceptions, e.g.\,HH\,1042).

\section{Summary}{\label{sec:sec4}
We have exploited the exceptionally bright X-shooter spectrum of the Herbig-Haro object HH\,1 to observationally derive Einstein $A-$coefficient ratios of bright [\feii] lines commonly observed 
in the spectra of nebular environments. These were derived from the observed intensity ratios of lines emitted from the same level, once properly de-reddened for the extinction toward the target. This latter was accurately determined by means of independent tracers, namely hydrogen recombination lines, atomic fine structure lines and H$_2$ ro-vibrational lines. We are able to give a meaningful range of $A-$coefficient ratios for seven pairs of [\feii] lines, which have been compared with the theoretical determinations available in the literature. Notably, none of these latter is consistent with {\it all} our determinations. The comparison of the intrinsic $\lambda$12570/$\lambda$16440 and $\lambda$13209/$\lambda$16440 intensity ratios derived in this work and a number of observations, shows
that our determinations are those that better agree with a large set of observations  although a significant number of literature data still lie to the right of the reddening line. Further observations are required 
to establish if  such discrepancy is due to instrumental effects, as we hypothesize.\\
In conclusion, although our analysis is limited to few bright lines, it represents a valuable effort to \emph{observationally} solve the problem of the determination of the $A-$coefficient ratios of [\feii] lines. These new determinations, which involve lines commonly observed in nebular environments, will be of great help in tracing the physical conditions (temperature, density, extinction) of the emitting gas.
\section{Acknowledgments}
We kindly thank A. Hibbert for providing the Deb \& Hibbert (2011) list of $A-$coefficients of Fe$^+$ lines.  
The ESO staff is acknowledged for support with the observations and the X-shooter pipeline. Financial support from INAF, under  PRIN2013 programme 'Disks, jets and the dawn of planets'
is also acknowledged.

%%%%%%%%%%%%%%%%%%%%%%%%%%%%%%%%%%%%%%%%%%%%%%%%%%%%%%%%%%%%%%%%%%%%%%%%%%%%%%%%%
\begin{deluxetable}{ccc}
\tablecaption{\label{tab:tab1} Fluxes of bright [\feii] lines.}
\tabletypesize{\small} 
\tablewidth{0pt}
\tablehead
{$\lambda_{vac}$&	Term                      & F$\pm\Delta$F \\
 (\AA)      &   Lower - Upper             & (10$^{-15}$erg s$^{-1}$cm$^{-2}$)}
\startdata
4289        & $a^6\!D_{9/2}-a^6\!S_{5/2}$ & 0.386$\pm$0.002\\
4360        & $a^6\!D_{7/2}-a^6\!S_{5/2}$ & 0.294$\pm$0.002\\
7174        & $a^4\!F_{7/2}-a^2\!G_{7/2}$ & 0.320$\pm$0.001\\
7390        & $a^4\!F_{5/2}-a^2\!G_{7/2}$ & 0.247$\pm$0.001\\ 
7639        & $a^6\!D_{7/2}-a^4\!P_{5/2}$ & 0.276$\pm$0.001\\ 
8619        & $a^4\!F_{9/2}-a^4\!P_{5/2}$ & 2.605$\pm$0.002\\
12570       & $a^6\!D_{9/2}-a^4\!D_{7/2}$ & 26.56$\pm$0.005\\
13209       & $a^6\!D_{7/2}-a^4\!D_{7/2}$ & 7.461$\pm$0.002\\
15339       & $a^4\!F_{9/2}-a^4\!D_{5/2}$ & 2.63$\pm$0.02\\
15999       & $a^4\!F_{7/2}-a^4\!D_{3/2}$ & 1.57$\pm$0.015\\
16440       & $a^4\!F_{9/2}-a^4\!D_{7/2}$ & 23.93$\pm$0.030\\
16773       & $a^4\!F_{7/2}-a^4\!D_{5/2}$ & 2.11$\pm$0.02\\
17116       & $a^4\!F_{5/2}-a^4\!D_{3/2}$ & 0.433$\pm$0.002\\
\enddata
\end{deluxetable}
%%%%%%%%%%%%%%%%%%%%%%%%%%%%%%%%%%%%%%%%%%%%%%%%%%%%%%%%%%%%%%%%%%%%%%%%%%%%%%%%%

\begin{deluxetable}{cccccc|cc}
\tablecaption{\label{tab:tab2} Comparison of $A-$coefficient ratios of bright [\feii] lines}
\tabletypesize{\footnotesize} 
\tablewidth{0pt}
\tablehead
{Ratio			    		    &   NS     & Q-SST   &   Q-HFR  &  DB          & SH        & \multicolumn{2}{c}{This work}   \\
                                & \multicolumn{5}{c|}{$A-$coefficient ratios}              & $A-$coefficient ratios   & Intrinsic intensity ratios$^a$ }
\startdata
$\lambda$12570/$\lambda$16440   &  1.04    & 0.79    &   0.90   & 1.04         &  1.13     & 0.84-0.92    & 1.11-1.20        \\
$\lambda$13209/$\lambda$16440   &  0.29    & 0.22    &   0.24   & 0.29         &  0.32     & 0.25-0.27    & 0.31-0.33        \\
$\lambda$15339/$\lambda$16773   &  1.26    & 1.25    &   1.26   & 1.25         &  1.07     & 1.12-1.18    & 1.22-1.27        \\
$\lambda$15995/$\lambda$17116   &  3.56    & 3.54    &   3.57   & 3.55         &  3.05     & 3.20-3.59    & 3.43-3.84        \\
$\lambda$7639/$\lambda$8619     &  0.13    & 0.19    &   0.23   & 0.43         &   -       & 0.09-0.10    & 0.10-0.12        \\
$\lambda$7174/$\lambda$7390     &    -     & 1.31    &   1.31   & 1.31         &   -       & 1.25-1.30    & 1.29-1.34      \\
$\lambda$4289/$\lambda$4360     &    -     & 1.34    &   1.35   &  -           &   -       & 1.25-1.32    & 1.27-1.34       \\
\enddata
\tablenotetext{a}{Intrinsic intensity ratios, computed multiplying the derived $A-$coefficient ratios times the vacuum frequencies ratios, taken from the 'The atomic line list, v2.04' (http://www.pa.uky.edu/~peter/atomic/).}
\tablecomments{References: NS: Nussbaumer \& Storey (1988); Q-SST: Quinet et al. (1996) - SuperStructure; Q-HFR:  Quinet et al. (1996) - Relativistic Hartree-Fock;  DB: Deb \& Hibbert (2011); SH: Smith \& Hartigan (2006).}\\
\end{deluxetable}

%%%%%%%%%%%%%%%%%%%%%%%%%%%%%%%%%%%%%%%%%%figure 1 %%%%%%%%%%%%%%%%%%%%%%%%%%%%%
\begin{figure*}
\includegraphics[width=15cm]{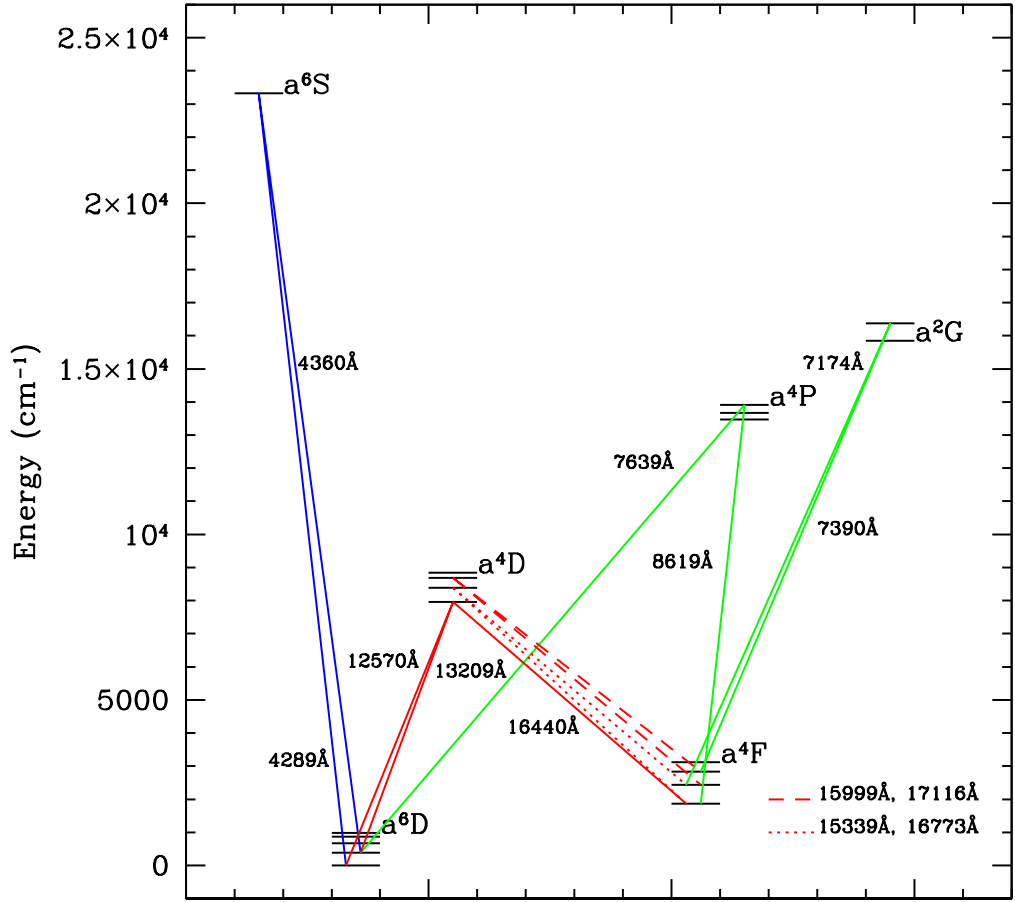}
\caption{Grotrian diagram of [\feii] with evidenced the lines discussed in this paper. UVB, VIS and NIR lines are depicted in blue, green, and red, respectively. \label{fig:fig1}}
\end{figure*}
%%%%%%%%%%%%%%%%%%%%%%%%%%%%%%%%%%%%%%%%%%%%%%%%%%%%%%%%%%%%%%%%%%%%%%%

%%%%%%%%%%%%%%%%%%%%%%%%%%%%%%%%%%%%%%%%%%figure 2 %%%%%%%%%%%%%%%%%%%%%%%%%%%%%
\begin{figure*}
\includegraphics[width=15cm]{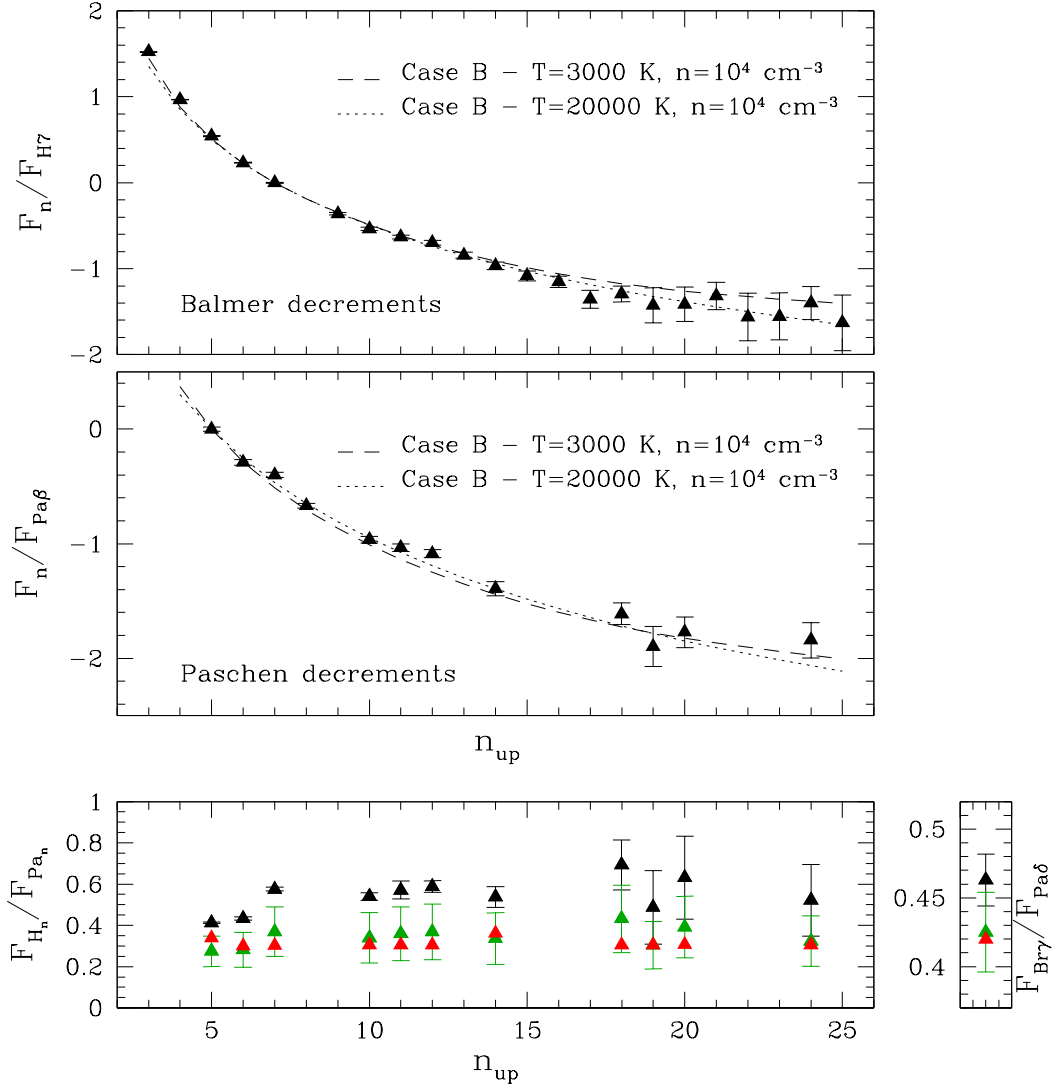}
\caption{Top and middle panels: Balmer and Paschen observed decrements fitted with case B recombination models. The best fits correspond to $n$=10$^4$ cm$^{-3}$ and 3\,000\,$\le\,T\,\le$\,20\,000 K, obtained by normalizing the Paschen decrements to the P$\beta$ and the Balmer decrements to the H7. 
In both panels we plot only the lines which are not blended with lines from other species and not contaminated by artifacts or sky line residuals in the spectral images. Bottom panel, left: ratios of Balmer and Paschen lines coming from the same upper level. Black symbols are 
the observed ratios, red symbols are the theoretical values and green symbols are the observed ratios once de-reddened for A$_V$=0.2$-$0.4 mag (estimated for R$_V$=3). Bottom panel, right: as in the left panel for the Br$\gamma$/Pa$\delta$
ratio. \label{fig:fig2}}
\end{figure*}
%%%%%%%%%%%%%%%%%%%%%%%%%%%%%%%%%%%%%%%%%%%%%%%%%%%%%%%%%%%%%%%%%%%%%%%

%%%%%%%%%%%%%%%%%%%%%%%%%%%%%%%%%%%%%%%%%%figure 3 %%%%%%%%%%%%%%%%%%%%%%%%%%%%%
\begin{figure*}
\includegraphics[width=16cm]{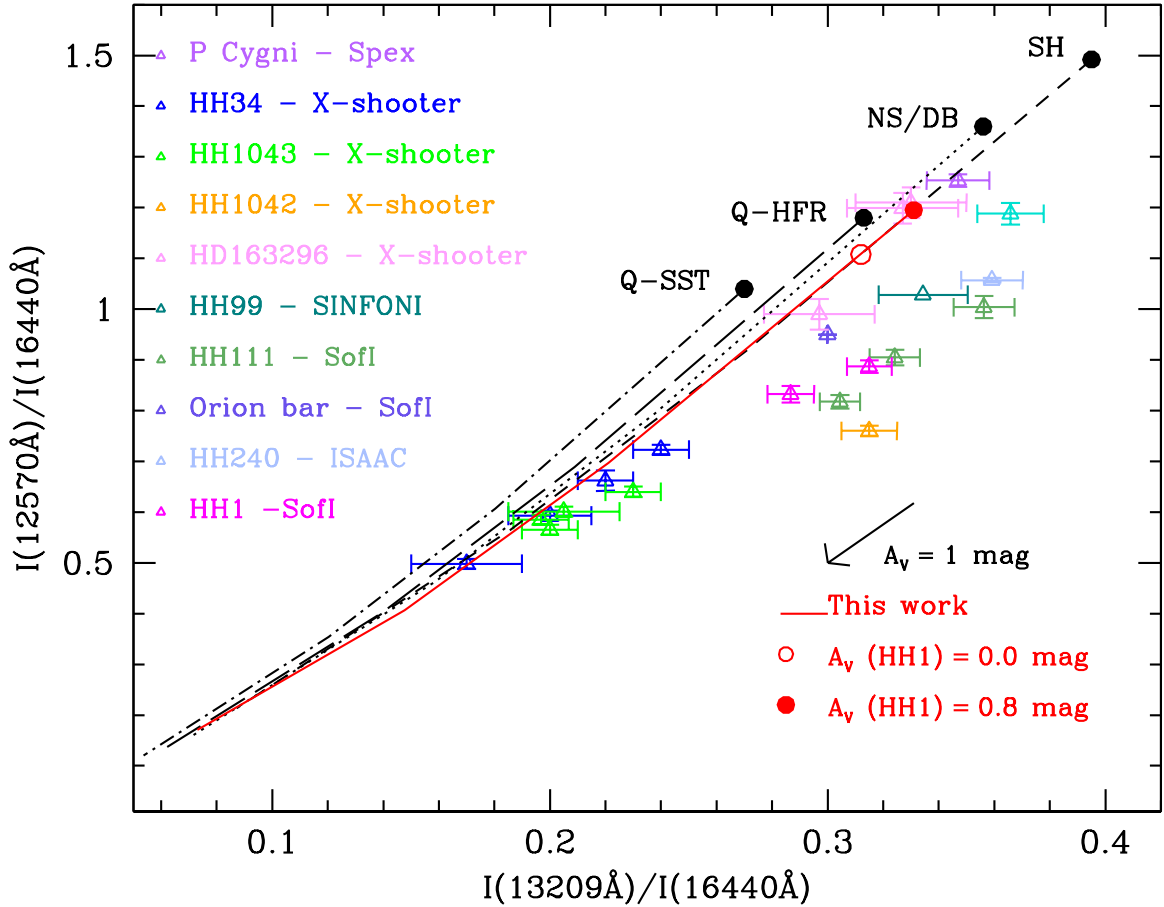}
\caption{I(12570\AA)/I(16440\AA) vs. I(13209\AA)/I(16440\AA) measured in bright objects (depicted with different colours, note that some objects have several data points, corresponding to different knots of emission). Intrinsic line ratios theoretically predicted (Q-SST = Quinet et al.
1996 - SuperStructure; Q-HFR = Quinet et al. 1996 - Relativistic Hartree-Fock; NS = Nussbaumer \& Storey 1988; DB = Deb \& Hibbert 2011), along with the empirical point by Smith \& Hartigan (2006, SH)
are labeled. The black curves represent 
the extinction law by Cardelli et al. (1988), applied to the theoretical points; in red is shown the same curve derived in the present work, which starts from the open or filled circle if A$_V$ toward HH\,1 is 0.0 or 0.8 mag.  The extincion vector corresponding to A$_V$\,=\,1 mag is depicted as well. Data from literature: HH111, HH240: Nisini et al. 2002; HH1: Nisini et al. 2005; P Cygni: Smith \& Hartigan 2006; HH99: Giannini et al. 2008; Orion bar: Walmsley et al. 2000; HH34: Nisini et al. 2014, in prep.; HH1042, HH1043: Ellerbroek et al. 2013; HD 163296: Ellerbroek et al. 2014. \label{fig:fig3}}
\end{figure*} 
%%%%%%%%%%%%%%%%%%%%%%%%%%%%%%%%%%%%%%%%%%%%%%%%%%%%%%%%%%%%%%%%%%%%%%%

\end{document}